\documentstyle[sprocl,psfig]{article}

\def\be{\begin{equation}}
\def\ee{\end{equation}}
\def\bea{\begin{eqnarray}}
\def\eea{\end{eqnarray}}

\begin{document}

\title{{\bf RANDOMNESS and COLLECTIVITY in NUCLEAR STRUCTURE: \\
Three theoretical puzzles}}
                                                       
\author{ Vladimir ZELEVINSKY and Alexander VOLYA}

\address{Department of Physics and Astronomy and\\
National Superconducting Cyclotron Laboratory,\\
East Lansing, MI 48824-1321 USA}


\maketitle\abstracts{We show and interpret three examples of nontrivial 
results obtained in numerical simulations of many-body systems: exponential
convergence of low-lying energy eigenvalues in the process of progressive
truncation of huge shell-model matrices, apparently ordered spectra generated by
random interactions, and regular behavior of complex many-body energies in a
system with single-particle orbitals in continuum. The possible practical
applications and new approaches are suggested.}

\section{Introduction}

As we discussed \cite{ald0} at the previous conference (S. Agata sui due Golfi,
1998), the ideas of quantum chaos significantly advance our understanding of 
many-body quantum systems. The new stage of this development is
related to the role of incoherent interactions between the
constituents. Apart from great theoretical interest, which extends to similar
problems in other finite many-body systems as atomic clusters, quantum dots and
atomic gases in traps, the progress in this direction would have practical
implications for the development of new approaches to the solution
of the quantum many-body problem. 

Instead of making an attempt to cover recent ideas in a systematic way, we show
three examples - puzzles which emerge from numerical
modeling of the nuclear many-body problem in restricted Hilbert space. 
In all three cases, the effect is very clear but its full understanding
requires significant theoretical efforts and is not completed until now
although below we give plausible explanations.

\section{Applying exponential convergence}

The dimensions of shell model spaces increase dramatically with the number of
single-particle orbitals included. This precludes the full shell model
diagonalization in practically interesting regions of the nuclear chart. Even
in the $fp$-shell one still awaits for the full calculation for all nuclei.
On the other hand, such a full solution would provide too much unphysical
information which
is not observable and, moreover, unstable with respect to small variations of
the interaction hamiltonian that is never known with high accuracy. In reality
we are interested in the properties of relatively few individual states
being satisfied with a statistical description for the rest of the spectrum.
The average properties of excited states were studied in the shell model
framework with the methods of statistical spectroscopy \cite{wong} and, with
direct diagonalization, in
relation to quantum chaos for atoms \cite{grib} and nuclei \cite{big}.
One can expect that a reasonable truncation of the shell model space 
could be possible when the influence of the remaining part of the basis is 
accounted for in some average sense.

\begin{figure}
\psfig{figure=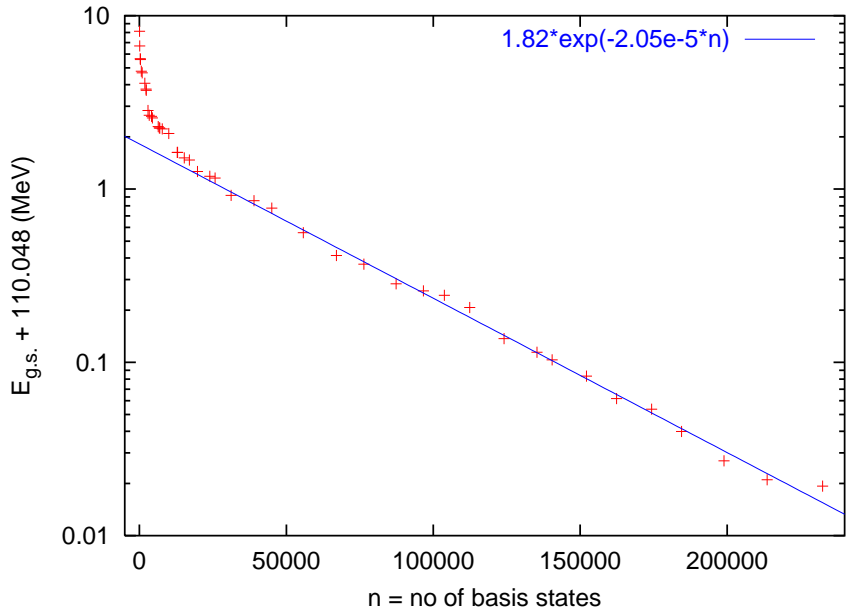,height=8cm,angle=-90}  
\caption{Ground state energy of $^{49}$Cr versus the dimension of trucated
shell model space (total $JT$-dimension is 232514).}
\end{figure}

Let us consider a behavior of the energy eigenvalue for the ground, or one
of low-lying, states, as a function of the matrix dimension used for
the exact diagonalization in the process of the progressive truncation of the
original huge shell model hamiltonian matrix. The generic picture is
shown in Fig. 1, taken from \cite{mihunhub}. After the initial
steep decline, the ground state energy of $^{49}$Cr 
monotonously descends to the exact value; the convergence is almost precisely
exponential. This is just one of many
available examples, both for the realistic shell model and for 
random Gaussian matrices. The property of exponential convergence seems to be
universal \cite{ecm}. The full analysis of validity can be performed for
tridiagonal matrices with a smooth change of matrix elements along the
diagonal.

The underlying physics is based on the saturation property of the energy
dispersion of simple, but spin-isospin ($JT$) projected, basis states 
\cite{big,fraz}. The centroid $\bar{E}_{k}$ and the width $\sigma_{k}$
of the basis state $|k\rangle$ can be found prior
to the diagonalization in terms of the matrix elements of the shell model
hamiltonian:
\begin{equation}
\bar{E}_{k}=H_{kk}, \quad \sigma_{k}^{2}=\sum_{k\neq l}H^{2}_{kl}.  \label{1}
\end{equation}
The widths $\sigma_{k}$ are nearly constant for all states of a given
$JT$-class, essentially because of the geometric chaoticity of angular momenta
coupling of individual particles. This justifies the recipe of statistical
spectroscopy \cite{wong} dealing with the centroid $\bar{E}$ and average width 
$\bar{\sigma}$ of each shell model partition. Expanding states $|k\rangle$ in
the eigenbasis $|\alpha\rangle$ of the hamiltonian, $|k\rangle=\sum_{\alpha}
C^{\alpha}_{k}|\alpha\rangle$, one can find the strength function
$F_{k}(E)=\sum_{\alpha}|C^{\alpha}_{k}|^{2}\delta(E-E_{\alpha})$ which also
reveals \cite{LBBZ} the saturation property as a function of $\bar{E}_{k}$. 
The generic shape of the strength function evolves, with the interaction 
strength increasing, from the standard Breit-Wigner to the Gaussian 
one \cite{fraz} . Among
various consequences of this evolution, one can mention \cite{lew}
the narrowing of the widths of multiple giant resonances,
$\Gamma_{n}\rightarrow \sqrt{n}\Gamma_{1}$. In the
strong coupling limit, the spreading width stabilizes on the level of
$\Gamma\approx 2\bar{\sigma}$. The strength 
fragmented to the states at an energy distance $>\Gamma$ should become less
and less important. Earlier we have suggested \cite{trun} the truncation 
scheme based on this idea. Now we can complement this with the exponential
extrapolation to the exact shell model result.

The detailed shell model analysis \cite{fraz} has established that the tails
of the strength function fall off almost pure
exponentially as a function of the energy distance from the centroid. This 
phenomenon, which reminds the exponential localization 
\cite{rammer} of electronic coordinate wave functions in disordered solids,
is also related to the nonexponential 
decay of nonstationary states at the early time stage. Based on this result, we
can assume that, in the inverted problem of the composition of the eigenstate
$|\alpha\rangle$, the remote basis states $|k\rangle$ give exponentially small
contributions. This is indeed seen in numerous examples. 

A practical application of the exponential convergence was recently
developed for the calculation of ground state
energies of nuclei in the $fp$-shell \cite{mihunhub}. One starts with the usual 
partition of the shell model space and calculates the average quantities
(\ref{1}). The configurations are ordered according to their energy centroids
$\bar{E}$ (this order might be very different from that in the particle-hole  
scheme). After that the diagonalization is performed with
progressive inclusion of new partitions in their ``natural" order.
At an energy distance of $(3-4)\bar{\sigma}$ from the original centroid, the 
exponential regime sets in \cite{mihunhub} for low-lying states, see Fig. 1, so
that extrapolating the energy dependence on the truncated dimension
$n$ as $E(n)=C+B\exp(-\gamma n)$, we can go to the limit of $n=N$, the full
space dimension.

Using the FPD6 interaction \cite{FPD6}, we calculated in this way ground state
energies, spins and isospins for all the lowest $|\Delta(N-Z)|$ nuclides from
$^{42}$Ca to $^{56}$Ni. Spins and isospins are reproduced correctly, except for
the case of $^{45}$Ti where the three levels with $J=3/2,5/2$ and 7/2 are
within 100 keV both in the experiment and in our calculation. As usual for such
calculations, the shell model energies (relative to the $^{40}$Ca) require 
Coulomb 
corrections \cite{langa} and additional monopole corrections \cite{mono}
taking into account the smooth evolution of the mean field with the valence
particle number. These corrections lead to the energy shift without changing
the wave functions. The resulting mean square deviation of ground state
energies from the data is 0.27 MeV. One can conclude that the exponential
convergence method is a powerful tool to be used in the shell model framework
for the cases when the full calculation is not feasible. The next step in this
direction should extend the method to the calculation of observables and
transition probabilities.

\section{Apparently ordered spectra from random interactions}

The second puzzle was formulated in the paper 
\cite{JBD} that triggered an extensive theoretical discussion \cite{disc,JBDT}.
Consider a finite shell-model space with a rotationally (or/and isospin)
invariant two-body interaction. The interaction is fully characterized by a set
of parameters $V_{LI}(j_{1}j_{2};j_{3}j_{4})$ corresponding to the scattering
$(j_{3},j_{4})\leftrightarrow (j_{1},j_{2})$ of the fermion
pair in the channel with 
total spin $L$ and isospin $I$ conserved in the process.
Let us randomly select these entries from a matrix ensemble which is 
more or less arbitrary but hermitian, real and symmetric with respect to the 
sign of the matrix elements $V_{LI}$.
If, for simplicity, the single-particle energies are kept degenerate, what will
be the distribution function of the quantum numbers of total spin $J$ and 
isospin $T$ of the many-body ground state generated by this ensemble? 

(i) The first idea coming to our mind is that 
any $JT$-set has a chance to have the lowest energy so that the resulting 
probability is merely 
determined by the number of available levels with given $J$ and $T$ in
the Hilbert space. For example, for one kind of particles, according to 
a traditional consideration of the Fermi-gas level density,
the total number ${\cal N}(J)$ of levels with given $J$ can be
estimated as ${\cal N}_{{\rm stat}}(J)\propto (2J+1)\exp[-J(J+1)/\Theta]$ 
where $\Theta$ is
related to the statistical moment inertia determined by the average value of 
$m^{2}$, the squared single-particle angular momentum projection, in the
available space. The maximum of the statistical distribution corresponds to
$2J+1=(2\Theta)^{1/2}$. 

\begin{figure}
\psfig{figure=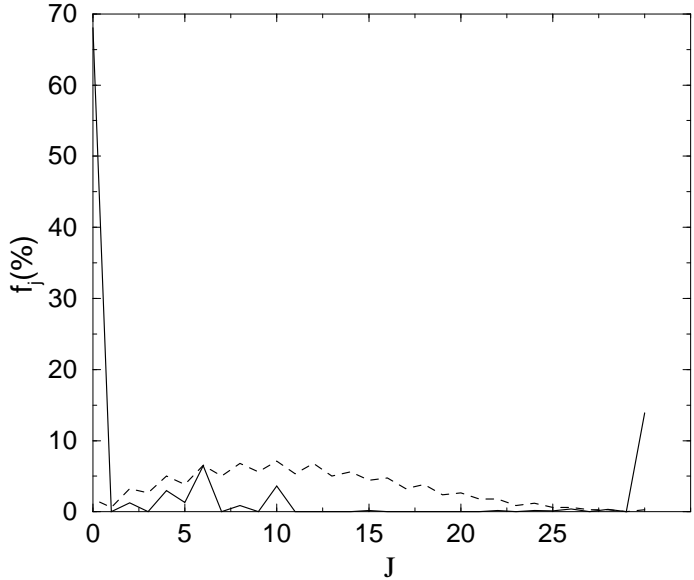,height=8.0cm}  
\caption{Fraction $f_{J}$ of ground states with spin $J$ in the uniform
ensemble of random two-body interactions for $N=6$ particles on the level
$j=15/2$; dashed line shows statistical multiplicities.}
\end{figure}

However, this idea turns out to be wrong. As shown in Ref. \cite{JBD} and
confirmed by many authors, the ground state spin is predominantly (typically
with probability exceeding 50\%) $J_{0}=0$, 
although the fraction of states of spin $J=0$ is
usually quite low. The existence of the effect is very 
robust and 
insensitive to the peculiarities of the ensemble. Its magnitude depends on
the choice of the ensemble and can exceed 90\%. The
preponderance of $J_{0}=0$ was found also in interacting
boson models \cite{boson}. In many cases the fraction of the
ground states with maximum possible spin, $J_{0}=J_{m}$, is also enhanced (the
statistical fraction of such states is very low; in a single-$j$ model, Fig. 2,
the state with $J=J_{m}$ is unique). We know that all even-even nuclei have
$J_{0}=0$. Is this fact originated from pairing forces as suggested by
the classics of the field \cite{maria,BMP}, or will the same pattern appear
with nearly any physically allowed interaction?

(ii) Statistical spectroscopy \cite{wong} 
teaches us to characterize
the general features of the spectra by the lowest moments of the
hamiltonian (centroid, width and so on). Comparing the statistical widths
$\sigma(J)$ of the subclasses 
with various values of spin $J$, we may expect that
if a class of states with given $J$ reveals the largest width (even if the
deep reason of that is still unclear) the states of this class will be
most probably the ground states of the system. However, for a majority of
ensembles, this conjecture fails. Fig. 3 shows the widths $\sigma_{J}$ 
in the single $j$-level space for the ensemble of matrix elements 
$V_{L}, L=0,2,...,2j-1$, uniformly distributed 
between $-1$ and +1 (a system of 6 identical particles). Although the ensemble 
leads to the dominance of $J_{0}=0$, the statistical width  
$\sigma_{J=0}$ is not maximum ; in some cases even $\sigma_{2}>
\sigma_{0}$. Moreover, to get a significant excess of the ground state
probability, the corresponding width should be considerably greater than others
which almost never happens.

\begin{figure}
\psfig{figure=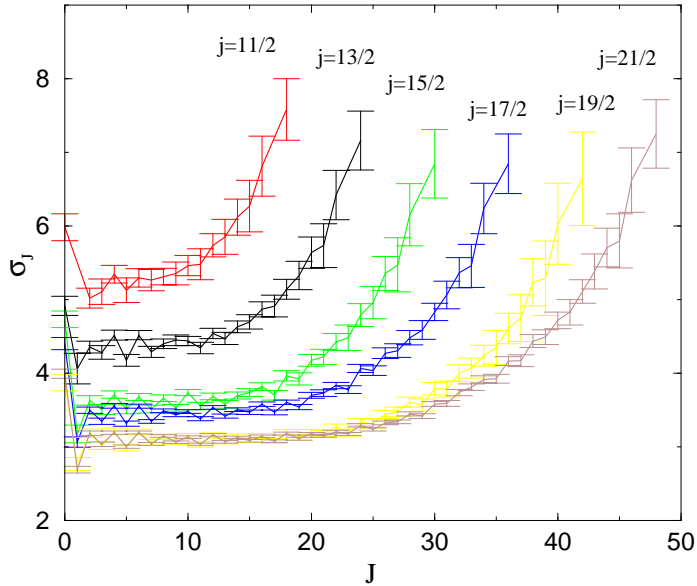,height=8.0cm}  
\caption{Widths $\sigma_{J}$ of the level density $\rho_{J}(E)$ in the
single-$j$ level shell model with random two-body interactions.}
\end{figure}

If the level density at energy $E$ for spin $J$ is $\rho_{J}(E)$,
the probability of finding a state of spin $J$ as a ground state can be 
formally defined as
\begin{equation}
f_{J}= \int_{-\infty}^{\infty} dE\,\left[-\frac{d}{dE}\chi_{J}(E)\right]
\prod_{J'\neq J}\chi_{J'}(E),                        \label{2}
\end{equation}
where, in terms of the level densities normalized to 1, $g_{J}(E)=\rho_{J}(E)/
{\cal N}_{J}$,
\begin{equation}
\chi_{J}(E)=\left(\int_{E}^{\infty}dE'\,g_{J}(E')\right)^{{\cal N}_{J}}.
                                                        \label{3}
\end{equation}
For uncorrelated densities $\rho_{J}$, eq. (\ref{3}) indeed prefers the class of
states with the greatest width. However, this conclusion is not valid since 
the densities are strongly correlated being determined by the same
interaction matrix elements. The task of calculating the many-point
correlation function of level densities is very hard.
 
(iii) Another alluring idea is that the dominance of $J_{0}=0$ is related to
the time-reversal invariant character of the random hamiltonian. If so, the
output could be different for a rotationally invariant, hermitian but complex 
hamiltonian. Physically this can be associated with the fact that the presence
of $J_{z}\neq 0$ acts as if time-reversal symmetry were broken by
selecting a sense of rotation; the corresponding quasi-Goldstone mode would be
rotation restoring symmetry by the transformation to another projection $J_{z}$.
This idea in the simplest form (introducing an imaginary part of the random
matrix element $V_{L}$) does not work \cite{disc} because the ensemble average
eliminates all imaginary terms along with the odd powers of $V_{L}$. Still, the
idea is promising if associated with the spontaneous symmetry breaking
which can be accomplished by the consideration of the body-fixed frame, see
below.

(iv) In the first paper \cite{JBD} on the subject, see also \cite{JBDT},
it was suggested that usual pairing correlations and the phonon 
collectivity emerge somehow from the random forces. This statement is correct
in a limited sense. Indeed, each realization of the two-body hamiltonian
in a many-level shell model space generates its own mean field. Then it is
possible to construct the superposition of the particle-hole operators of a
given multipolarity (a generalized phonon) which would maximize the coherence
and give an enhanced transition probability. In a similar way one can look for
the specially selected generalized seniority operator \cite{JBDT}
to  enhance the pair
transfer processes. Those operators are different in different copies of
the ensemble. A comparison with a standard paired state shows that its overlap
with the ground states of random interactions is quite small, both in a single
$j$-case \cite{rand} and in a realistic shell model \cite{versus}. The phonon
collectivity with a fixed multipole operator is also absent.

(v) Currently the only plausible explanation of the preference of $J_{0}=0$
ground states in randomly interacting systems is based \cite{rand,shed} on the
idea of geometric chaoticity. With a random interaction we do not expect any
specific shape of the mean field to be singled out. The wave functions are
very complicated combinations of shell model basis states. Therefore a
statistical approach seems to be suitable which looks for the single-particle
density matrix with maximum entropy
under constraints of fixed particle number $N$ and total spin (isospin). 
In a single $j$-model, the density matrix is diagonal for the aligned state
with the total projection $M=J$ (analog of the body-fixed frame). Its
eigenvalues give the single-particle occupation numbers $n_{m}=[\exp(\gamma m
-\mu)+1]^{-1}$, where the Lagrange multipliers of chemical potential $\mu$ and
cranking frequency $\gamma$ fix the average values of $N$ and $M$. The
expectation value of the total hamiltonian calculated with such occupation
numbers gives the simple approximation for the average
yrast line. Depending on the
sign of the effective moment of inertia for the given set of $V_{L}$, this
leads to $J_{0}=0$ or $J_{0}=J_{m}$, a normal or inverted band,
respectively. Thus we come to a trivial geometric mechanism of the preference
for the edge values of the ground state spin. With some improvements, one can
reproduce average empirical results. 

The energy values estimated with this statistical approach correlate well
with the exact numerical values although there is a small systematic
discrepancy \cite{rand} in the probability $f_{J}$ for small $J$ (the agreement
is almost perfect for high $J$ including $J_{m}$). One source of the deviation
is in the approximation of the expectation value $\langle\hat{n}_{m}
\hat{n}_{m'}\rangle$ 
by the product of two statistical average values $n_{m}n_{m'}$.
In a given wave function the presence of the mean field makes the occupancies  
slightly dynamically correlated. Apart from that, there is indeed some
coherence generated apparently by the off-diagonal matrix elements of the
interaction in higher (even) orders. This brings in a small excess of 
the overlap of the ``random" ground state with the paired one compared to a
pure statistical (in the sense of the random matrix theory) estimate
\cite{versus}. By the same reason, the percentage $f_{0}$ of the ground states 
with $J_{0}=0$ increases when going from the single-$j$ case to a more
realistic shell model scheme, especially to the set of many spin 1/2 levels,
when the role of the off-diagonal pair transfer
elements is the most important.
This can be seen also in the fact of the overwhelming percentage of the lowest
isospin in the ground states. Another example is given \cite{versus} by
the specific ensemble which includes only random pairing matrix elements;
the sign-independent effect of the off-diagonal pair transfers leads to the
percentage $f_{0}>90\%$.

Such dynamical effects are still not fully understood albeit they may 
be the most interesting and essential for many-body
physics in finite quantum systems. The ideas of solving exactly the coherent
parts of the interaction (for example, pairing \cite{EP}) and accounting in a 
statistical spirit for incoherent collision-like processes are in the air
promising a new interesting development in the near future.

\section{Approaching the continuum}

The third puzzle comes from an attempt to generalize the shell-model approach
for loosely bound or unbound nuclei where the entire dynamics take place on the
edge of or already within the continuum. 
A progress in this direction is essential
both for nuclear physics far from stability and for astrophysics. Since the
various versions of the shell model with the discrete spectrum work 
exceedingly well for stable nuclei, it is tempting to consider a realistic mean
field where some single-particle orbitals are resonances in the continuum 
\cite{liotta} and include the residual interaction in order to obtain the
observable positions and widths of the many-body states.

\begin{figure}
\psfig{figure=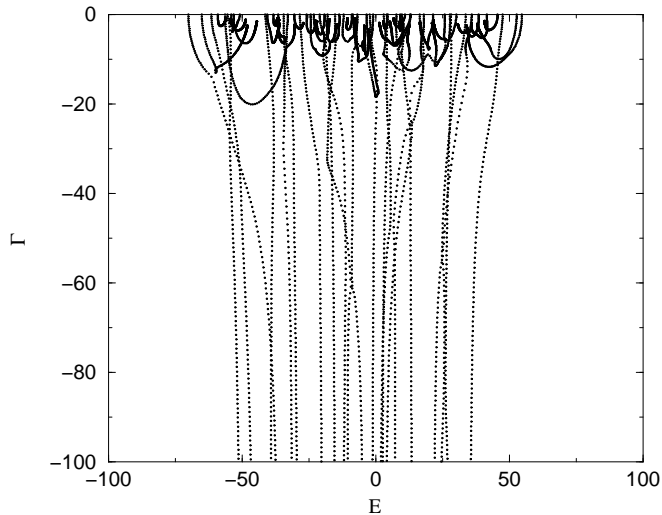,height=7.0cm}  
\caption{Dynamics of complex energies $E-(1/2)\Gamma$ for a system of 3
fermions on 8 equidistant orbitals interacting via random interaction; 
the upper orbital has a single-particle width $\gamma$, and the resonances move
as $\gamma$ increases.}
\end{figure}

One example of what happens in such a problem is given by Fig. 4. We assume
that the single-particle levels $\epsilon_{\nu}$ have some decay width; for 
simplicity we attribute here a significant width to one upper level 
shifting its energy to the complex plane,
$\epsilon\rightarrow \epsilon-(i/2)\gamma$. Let us switch on a two-body
interaction with real random matrix elements $V$ (here we do not observe any
conservation laws so that all pairs of orbitals are mutually coupled) and
find the eigenvalues of the many-body system. Fig. 4 corresponds to the case of
3 particles on 8 equidistant orbitals; the trajectories of the complex energies
${\cal E}=E_-(i/2)\Gamma$ 
are shown as functions of increasing instability $\gamma$ of
the upper orbital. Instead of complete chaos, we see a more and more regular 
pattern as $\gamma$ increases. Interestingly enough, 21 energies
move almost parallel to each other into the complex plane whereas 35 states
have a very small width.

It is easy to understand this dynamics of complex energies (or poles of the
scattering matrix). The total number of many-body states in this truncated 
space is $8!/(3!4!)=56$. In the limit of large $\gamma$, any state which has 
the upper orbital filled, even with a low probability, will decay very fast.
The number of such states corresponds to a number of combinations of the
remaining particles within the rest of space, $7!/(2!5!)=21$. The increasing
original decay width is distributed over the many-body states, since the
imaginary part of the trace of the hamiltonian is preserved. As $\gamma$ grows, 
the ``self-organization" occurs: fast and slow decaying states are separated in
time. In a reaction populating the system, one would see two distinct time
scales, corresponding to direct and compound processes. Thus, coupling to
the continuum can bring order in a system governed by a random hamiltonian. 

The physics we are looking at here was extensively discussed earlier
from a different viewpoint \cite{Krot,sok,sokann}. In that approach one
starts with the set of many-body states $|\alpha\rangle$ formed by a normal
hermitian interaction hamiltonian $H$. The coupling to the continuum is given by
the antihermitian part of the effective energy-dependent hamiltonian,
\begin{equation}
W_{\alpha\beta}=\sum_{c}A^{c}_{\alpha}A^{c\ast}_{\beta}. \label{5}
\end{equation}
Here the sum runs over all decay channels $c$ that are {\sl open}  
at a given energy,
and $A^{c}_{\alpha}$ is the decay vertex of an intrinsic state
$|\alpha\rangle$ into a channel $c$.
The factorizable form of eq. (\ref{5}) comes from the on-shell 
contribution of the effective propagator for intrinsic states coupled through
the continuum \cite{mahau} and unitarity requirements. The 
complex eigenvalues of the total effective hamiltonian ${\cal H}=H-(i/2)W$
give the resonance energies. In the weak continuum coupling regime, $W$ is a
perturbation providing narrow resonance widths. As this coupling becomes
strong, a phase transition occurs to the overlap regime with Ericson
fluctuations of cross sections and separation of the time scales. A number of
states (equal to the number of open channels) gives rise to broad
short-lived resonances absorbing the lion's share of the total width
while the remaining compound states become long-lived and reveal
internal thermalization and equilibration. This phenomenon, being
an analog of the Dicke superadiance in optics \cite{dicke}, has
interesting applications to physics of giant resonances \cite{sokrot}.

In the example above, strictly speaking, all many-body states are
nonstationary. In reality there exists a set of threshold energies $E^{(c)}$
determined by the $Q$-values of a reaction in channel $c$. The amplitudes
$A^{c}_{\alpha}$ depend on running energy $E$ and 
have a branching point at threshold \cite{sokann}, for example in the case
of decay into an $s$-wave in the continuum, $A^{c}_{\alpha}\propto
(E-E^{(c)})^{1/2}$. Therefore, in principle, the hybrid approach combining
the shell model with the effective nonhermitian hamiltonian allows for a
self-consistent calculation of discrete levels, resonances and reaction cross
sections. Being technically very difficult, this problem is of vital importance
for physics of weakly bound nuclei. Far away of
thresholds, the method of a 
nonhermitian hamiltonian was used \cite{simple} for 
the microscopic derivation and analysis of the kinetics of resonance population
and decay. Two phenomena, the loss of the 
collective strength \cite{gaard}, and the restoration of isospin purity
at high excitation energy \cite{kurt} naturally follow from this consideration.

\section{Conclusion}

Three ``puzzles" briefly discussed above show significant gaps in current theory
of nuclear structure and reactions. The blunt diagonalization of huge
shell-model matrices cannot be an optimal way of solving the nuclear many-body
problem. Even randomly taken but geometrically correct interactions generate
some features of observed regular spectra. The presence of continuum aligns
the intrinsic states along the new ``axis" related to their ability to
decay. Those are just particular examples of new avenues which should be
actively studied. 

\section*{Acknowledgments}
 
The authors acknowledge support from the NSF grants 96-05207 and 0070911.

\end{document}